# Performance Limits for Field Effect Transistors as Terahertz Detectors

V. Yu. Kachorovskii[1,2], S. L. Rumyantsev[1,2,3], W.Knap[3], and M. Shur[1]

[1]*Rensselaer Polytechnic Institute, 110, 8th street, Troy, NY, 12180, USA.*
[2]*A.F. Ioffe Physical-Technical Institute, 26 Polytechnicheskaya Street, St. Petersburg, 194021, Russia*
[3]*Laboratoire Charles Coulomb &TERALAB, Université Montpellier 2 & CNRS, 34950 Montpellie, France*

We present estimates of the performance limits of terahertz detectors based on the field effect transistors (FET) in the regime of broadband detection. The maximal responsivity is predicted for short-channel FETs in the subthreshold regime. We also calculate the conversion efficiency $Q$ of the device defined as the ratio of the power dissipated by radiation-induced dc current to the THz dissipated power. We show that $Q$ has an absolute maximum as a function of two variables: the power and the frequency of the incoming radiation. The maximal value of $Q$ is on the order of 10%.

## I. INTRODUCTION

Recent progress in terahertz (THz) plasmonic technology[1-21] has promise for applications in homeland security, biomedical imaging, radio astronomy, industrial controls, and short range covert and space communications. The plasmonic technology uses field-effect transistors (FETs) as compact and tunable emitters, detectors, and modulators of THz radiation. The nonlinear properties of the FET channel may be used for detection of THz radiation.[2] Due to the nonlinearity, the potential and density oscillations induced by the incoming THz wave with a frequency $\omega$ are rectified into a dc voltage drop across the FET channel $\Delta U$ (the detector response). For the resonant case, $\omega_0\tau \gg 1$ (here $\omega_o$ is the fundamental plasma frequency and $\tau$ is the momentum scattering time), the plasma oscillations with a high quality factor are excited in the FET channel, when the signal frequency $\omega$ is close to $\omega_0$ and its harmonics. In the opposite limit, $\omega_0\tau \ll 1$ (the so-called broadband detection regime), intensive scattering prevents excitation of plasma waves and incoming THz radiation leads only to overdamped plasma oscillations. In this paper, we only consider the broadband detection, which is a typical regime of operation at room temperature and above.

The plasmonic approach[1,2] to THz electronics was strongly supported by experimental observation of both resonant and broadband detection as well as emission of sub-THz and THz radiation by nanoscale FETs. Though FET-based THz emitters are still rather weak, the THz detectors already show good performance: they are tunable[3-21] by varying the gate voltage and the drain current, and demonstrate fast response time[13,21] and relatively low noise equivalent power up to room temperature[10,12]. Recent works has demonstrated responsivity in a FET up to 5 kV/W for conventional detectors,[18] and 90 kV/W for detectors with amplifiers.[14,18] These numbers, however, do not characterize the intrinsic device performance, since the responsivity was calculated by dividing the THz induced voltage drop across the FET over the THz power on the antenna. Actually, the power applied to the FET is smaller due to inevitable losses in the interconnects and is affected by antenna structure. This paper analyses the intrinsic performance of plasmonic detectors by considering the response to the THz power, $P_\omega$, absorbed in the detector itself.

For small signals, the response, $\Delta U$, is proportional to $(U_a)^2$, where $U_a$ is the amplitude of the THz signal applied between gate and source.[2] Actually, this amplitude is unknown and usually used as a fitting parameter, while comparing theory with experimental data. As shown below, the power $P_\omega$, dissipated in the device at the THz frequency, is also proportional to $(U_a)^2$. Therefore, for small signals, the *voltage* and *current responsivities*, defined as

$$R_U = \Delta U / P_\omega, \qquad R_I = \Delta U / rP_\omega \qquad (1)$$

do not depend on $U_a$, being only functions of the device and materials parameters, signal frequency, and device bias. Here $r$ is the FET resistance.

We consider also another figure of merit - *conversion efficiency,* which is the ratio of the dc power at the detector output to the THz dissipated power:

$$Q = P_{dc} / P_\omega \qquad (2)$$

The maximal power at the load is obtained when the load resistance is equal to the resistance of the device, $r$, so that

$$P_{dc} = (\Delta U)^2 / r. \qquad (3)$$

In contrast to $R_U$ and $R_I$, conversion efficiency depends on the radiation power and, as we demonstrate below, has a maximum for a moderate signal levels.

The approach developed in this work is similar to the analysis of internal quantum efficiency often used to compare different visible and ultraviolet radiation detectors.

## II. ABOVE THRESHOLD REGIME

In this section we assume that the gate-to-source voltage swing, $U_g = U_{gs} - U_t$ is positive and large compared with $U_T$ (here $U_{gs}$ is the gate-to-source voltage, $U_T$ is the thermal voltage, and $U_t$ is the threshold voltage). The local voltage swing between gate and channel, $U$, obeys the following equation[2,11]

$$\partial U / \partial t + (\mu/2) \partial^2 U^2 / \partial x^2 = 0 \quad (4)$$

Voltage $U$ controls the local electron concentration, $n = CU/e$, where $C$ is the capacitance per unit area. We choose the following boundary conditions[1,2]

$$U\big|_{x=0} = U_g + U_a \cos \omega t, \quad \partial U / \partial x \big|_{x=L} = 0, \quad (5)$$

(here $L$ is the gate length) and search for the solution of Eq. (4) in the form of the Fourier expansion

$$U = U_0 + (U_1 e^{-i\omega t} + U_1^* e^{i\omega t})/2 + \ldots$$

For low input signals ($U_a \ll U_g$), one could linearize Eq. (4) neglecting higher harmonics:

$$-i(\omega/\mu) U_1 + \partial^2 (U_1 U_0) / \partial x^2 = 0, \quad (6)$$
$$U_1(0) = U_a, \partial U_1 / \partial x \big|_{x=L} = 0.$$

For $U_a \ll U_g$, one may assume that $U_0 \approx U_g = const$ in Eq. (6). The solution of Eq. (6) is

$$U = U_g + U_a \{ e^{-i\omega t} \cosh[q(x-L)] / 2\cosh(qL) + h.c. \}, \quad (7)$$

where $q = (1+i)/L_0$ and

$$L_0 = \sqrt{2\mu U_g / \omega}, \quad (8)$$

is the characteristic decay length.

Let us consider first the case of a *long sample*, $L \gg L_0$. In this case, we find from Eq. (7): $U_1 \approx U_a e^{(i-1)x/L_0}$. Neglecting the spatial dependence of $U_0$, we arrive to the following solution of Eq. (14):

$$U = U_g + U_a \cos(\omega t - x/L_0) e^{-x/L_0}. \quad (9)$$

Let us now calculate the power dissipated in the FET channel per unit width at the THz frequency

$$P_\omega = \left\langle \int_0^L jE\, dx \right\rangle_t = \left\langle \int_0^L \mu CU \left( \frac{\partial U}{\partial x} \right)^2 dx \right\rangle_t. \quad (10)$$

Here $\langle \cdots \rangle_t$ stands for time averaging over period $2\pi/\omega$ and we used the following equation for the local current in the channel: $j = \mu CU$. A simple calculation yields

$$P_\omega = \sigma U_a^2 / 2L_0, \quad (11)$$

where $\sigma = e\mu n = \mu C U_g$ is the channel conductivity. For a long sample, the above threshold non-resonant response is given by[2]

$$\Delta U = U_a^2 / 4U_g, \quad (12)$$

Using Eqs. (1), (11), (12) and expression for resistance per unit width, $r = L/\sigma$, we get

$$R_U = L_0 / 2\sigma U_g, \quad R_I = L_0 / 2L U_g, \quad (13)$$

Substituting $L_0$ from Eq. (8) one finds

$$R_U = U_g^{-3/2} (2\mu\omega)^{-1/2} C^{-1}, R_I = L^{-1} \mu^{1/2} (2U_g \omega)^{-1/2}. \quad (14)$$

Both $R_U$ and $R_I$ decrease with $\omega$ as $1/\sqrt{\omega}$. However, their mobility dependence is different: $R_U$ decreases with $\mu$ while $R_I$ increases.

Next, we discuss what happens in the *short sample*, where $L \ll L_0$. It is well known that in this case the response is much smaller compared to the case of a long sample:[2]

$$\Delta U = (U_a^2 / 6U_g)(L/L_0)^4. \quad (15)$$

However, the responsivity, turns out to be maximal for this case. Indeed, calculating $P_\omega$ from Eqs. (7) and (10), we get

$$P_\omega = (2\sigma U_a^2 / 3L)(L/L_0)^4. \quad (16)$$

Equation (16) implies that the dissipated power also decreases with $L_0$ in the same way: $\Delta U \sim 1/L_0^4$. The voltage and current responsivities obey

$$R_U = L/4\sigma U_g, \quad R_I = 1/4U_g. \quad (17)$$

One might also derive equations valid for arbitrary relation between $L_0$ and $L$:

$$P_\omega = \frac{\sigma U_a^2}{2L_0} \frac{\sinh(2L/L_0) - \sin(2L/L_0)}{\cosh(2L/L_0) + \cos(2L/L_0)}, \quad (18)$$

$$R_U = L_{eff} / 2\sigma U_g, \quad R_I = L_{eff} / 2L U_g. \quad (19)$$

Here

$$L_{eff} = L_0 \frac{\cosh(2L/L_0) + \cos(2L/L_0) - 2}{\sinh(2L/L_0) - \sin(2L/L_0)}$$
$$= \begin{cases} L_0, \text{ for } L_0 \ll L, \\ L/2, \text{ for } L_0 \gg L. \end{cases} \quad (20)$$

### III. BELOW THRESHOLD REGIME

The above calculations can be easily generalized for a FET operating below threshold. In the subthreshold regime, $U_g < 0$ and $|U_g| \gg U_T$. Here $U_T = \eta T / e$ is the thermal voltage and $\eta$ is subthreshold slope (sometimes called as ideality factor).

One may show that Eqs. (18) and (19) still hold with the replacement of $U_g$ with $U_T$:

$$R_U = L_{eff} / 2\sigma U_T, \quad R_I = L_{eff} / 2LU_T. \quad (21)$$

The effective length, $L_{eff}$, is given by Eq. (20), where $L_0$ is obtained from Eq. (8) by replacement $U_g \to U_T$:

$$L_0 = \sqrt{2\mu U_T / \omega} \quad (22)$$

Another essential difference from the above threshold case is that the conductivity entering denominator of $R_U$ is now exponentially small

$$\sigma = \mu C U_T \exp(-e|U_g|/\eta T). \quad (23)$$

This, in turn, means that $R_U$ is exponentially large below threshold. On the other hand, $R_I$, which is obtained from $R_U$ by dividing over $r$, does not contain such an exponential factor. The maximal value of current responsivity is obtained below-threshold for short transistors ($L_0 \gg L$):

$$R_I = 1/4U_T \quad (24)$$

This value is on the order of magnitude of the maximal current responsivity for Schottky diode.[22]

### IV. CONVERSION EFFICIENCY

The results presented above allow us to calculate the conversion efficiency. Below threshold we get:

$$Q = \beta U_a^2 / U_T^2. \quad (25)$$

Here

$$\beta = \frac{L_0}{8L} \frac{[\cosh(2L/L_0) + \cos(2L/L_0) - 2]^2}{\sinh(2L/L_0) - \sin(2L/L_0)} \times \frac{1}{\cosh(2L/L_0) + \cos(2L/L_0)}, \quad (26)$$

is the conversion coefficient, which is the function of $L_0/L$ only (see Fig. 1). We see that the maximal value of $\beta$ is about 0.05. In Fig. 2, we compare $\beta$ for GaN, InGaAs, and Si plasmonic THz FET detectors. As seen, the short channel (22 nm) Si FETs reach optimum performance close to 1 THz. The 22 nm InGaAs FETs exhibit superior performance at frequencies as high as 10 THz. GaN-based FETs are best suited for frequencies on the order of a few THz (with additional advantages of the superior dynamic range, which will be discussed elsewhere).

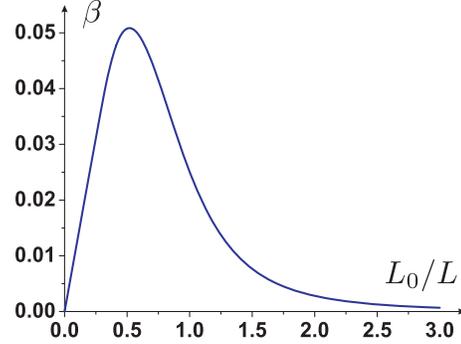

FIG.1. *Conversion coefficient $\beta$ as a function of $L_0/L$.*

### V. LARGE SIGNAL REGIME

Recent paper [20] demonstrated that the growth of response with $U_a$ slows down with increasing radiation power. In the most interesting subthreshold case, where response is maximal, one obtains $\Delta U \propto U_a$ for $U_a \gg U_T$ and $L_0 \ll L$, in contrast to $\Delta U \propto U_a^2$ for $U_a \ll U_T$.

Using the technique developed in Ref. [20] one may calculate responsivity for arbitrary relation between $U_a$ and $U_T$. Below we present the results for subthreshold regime. One may show that Eqs. (21) are still valid with $L_{eff}$ given by

$$L_{eff} = \begin{cases} L_0 \dfrac{2 \ln [I_0(U_a/U_T)]}{(U_a/U_T) I_1(U_a/U_T)}, & \text{for } L_0 \ll L, \\ \dfrac{L}{4} \dfrac{U_a/U_T}{I_1(U_a/U_T)}, & \text{for } L_0 \gg L. \end{cases}$$
(27)

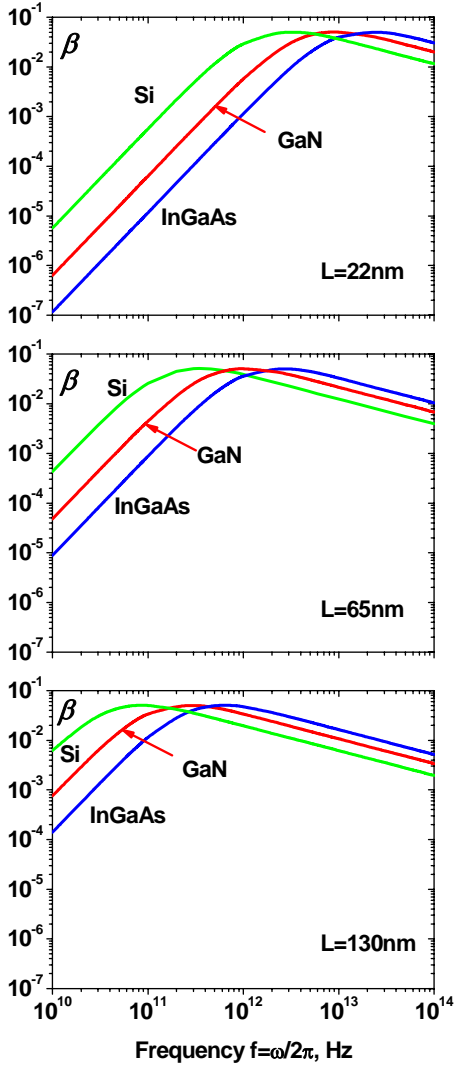

FIG.2. *Conversion coefficient β for different materials at different gate lengths as a function of frequency,* $f = \omega/2\pi$.

Here $I_0$ and $I_1$ are the Bessel function of imaginary argument. One can easily check that in the limit $U_a \to 0$, Eq. (27) reproduces asymptotics of $L_{eff}$ [see Eq. (20)] for long and short samples, respectively. From Eqs. (21) and (27), we find that the responsivity decreases with $U_a$. For example, responsivity in the short-channel FET ($L_0 \gg L$)

$$R_I = \frac{1}{8U_T} \frac{U_a/U_T}{I_1(U_a/U_T)} \quad (28)$$

coincides with Eq. (24) for $U_a \ll U_T$, and decays exponentially in the opposite case $U_a \gg U_T$:

$$R_I = (1/8U_T)\sqrt{2\pi}\,(U_a/U_T)^{3/2}\exp(-U_a/U_T). \quad (29)$$

Physically, the exponential decay is due to sharp dependence of electron concentration and, as a consequence, of the dissipated power on the gate-to-source swing. In contrast, response, $\Delta U$, is limited by $U_a$ and can not increase exponentially.

The above equations allow us to calculate the conversion efficiency. In particular, for $L_0 \ll L$, we obtain

$$Q = 2\beta^* L_0/L, \quad (30)$$

where

$$\beta^* = \frac{U_T}{2U_a}\frac{\ln^2\left[I_0(U_a/U_T)\right]}{I_1(U_a/U_T)}. \quad (31)$$

The dependence of $\beta^*$ on $U_a/U_T$ is plotted in Fig. 3.

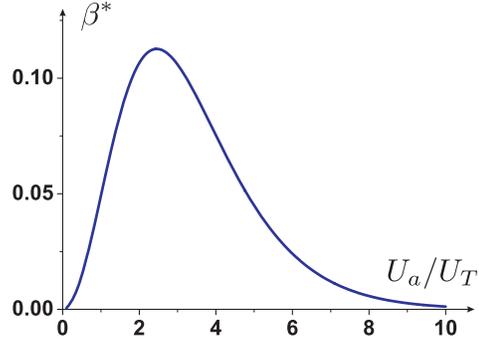

FIG.3. *Dependence of* $\beta^*$ *on* $U_a/U_T$.

Analyzing Eqs. (25), (26), (30) and (31) one comes to the conclusion that $Q$ has *the absolute maximum* as a function of two variables: the frequency (included into the factor $L_0/L$) and the power of the signal (encoded in the factor $U_a/U_T$). The maximum is achieved for $L_0/L \approx 0.5$ and $U_a/U_T \approx 2$. The maximal value of $Q$ is slightly higher than 0.1. At the point corresponding to the maximum, we have [see Eq. (18)]: $P_\omega \approx 2\sigma U_T^2/L_0$, where $L_0$ and $\sigma$ are given by Eqs. (22) and (23). Hence, for a given power $P_\omega$, the highest conversion efficiency (about 10%) is achieved for[23]

$$L \approx 2L_0 \approx 2\sigma U_T^2 / P_\omega. \qquad (32)$$

It is worth noting that the ac voltage applied to the devices automatically satisfies the optimal condition $U_a \approx 2U_T$ provided that $L$ and $L_0$ are chosen according to Eq. (32). We also stress that optimal conversion corresponds to intermediate strength of the signals.

The above calculations characterize the intrinsic performance of the FET itself. The overall optimization of the device performance also requires (see discussion in Refs [14,19] the matching of the device input *ac* impedance, $Z_\omega = (1+i)L_0 \coth[(1-i)L/L_0]/2\sigma W$, to the impedance of the external antennas and/or amplifiers (here $W$ is the FET width). Such optimization will be discussed elsewhere.

To conclude, we have presented calculation of the maximal achievable responsivity and conversion efficiency of FETs operating at THz frequencies. We showed that the conversion efficiency of FET has the absolute maximum (on the order of 10%) as a function of the device parameters.

This work was supported by ANR project "WITH" and by CNRS and GDR-I project "Semiconductor sources and detectors of THz frequencies" and by the US – French initiative "PUF". The work at University of Montpellier was supported by the Scientifique Interest Groupement GIS –TERALAB. The work at RPI was supported by the by Army Research Laboratory under ARL MSME Alliance. The work at Ioffe Institute was supported by RFBR and Programs of the RAS.